# Unexpected Tuning of the Anomalous Hall Effect in Altermagnetic MnTe thin films

Sara Bey[1], Shelby S. Fields[2], Nicholas G. Combs[2], Bence G. Márkus[1,3], Dávid Beke[1,3], Jiashu Wang[1], Anton V. Ievlev[4], Maksym Zhukovskyi[5], Tatyana Orlova[5], László Forró [1,3], Steven P. Bennett[2], Xinyu Liu[1], Badih A. Assaf[1]

[1] Department of Physics and Astronomy, University of Notre Dame, Notre Dame IN, 46556, USA.
[2] Materials Science and Technology Division, U.S. Naval Research Laboratory, Washington DC, 20375, USA
[3] Stavropoulos Center for Quantum Matter, University of Notre Dame, Notre Dame IN, 46556, USA.
[4] Center for Nanophase Materials Science, Oak Ridge National Laboratory, Oak Ridge, TN, 37831, USA
[5] Notre Dame Integrated Imaging Facility, University of Notre Dame, Notre Dame, IN, 46556, USA

**Abstract.** The discovery of an anomalous Hall effect (AHE) sensitive to the magnetic state of antiferromagnets can trigger a new era of spintronics, if materials that host a tunable and strong AHE are identified. Altermagnets are a new class of materials that can under certain conditions manifest a strong AHE, without having a net magnetization. But, the ability to control their AHE is still lacking. In this study, we demonstrate that the AHE in altermagnetic α-MnTe grown on GaAs(111) substrates can be "written on-demand" by cooling the material under an in-plane magnetic field. The magnetic field controls the strength and the coercivity of the AHE. Remarkably, this control is unique to α-MnTe grown on GaAs and is absent in α-MnTe grown on SrF2. The tunability that we reveal challenges our current understanding of the symmetry-allowed AHE in this material and opens new possibilities for the design of altermagnetic spintronic devices.

**Introduction**

Ferromagnets (FM) and antiferromagnets (AFM) are the two established classes of collinear magnetic materials. In FMs, a spontaneous magnetization is present, due to the parallel alignment of microscopic atomic spins below the Curie temperature. The broken time reversal symmetry (TRS) and spin polarization in the electronic band structure enables phenomena such as the anomalous Hall effect (AHE), a magneto-optical Kerr effect (MOKE), X-ray magnetic circular dichroism (XMCD), and polarized spin currents. In collinear AFMs, the lattice is divided into two sublattices connected by translation or inversion symmetry whose moments fully compensate each other, resulting in zero net magnetization, and zero spin polarization in the band structure (*1*). Hence, phenomena that result from broken TRS and spin polarization are not expected in AFMs. Now, a third class of materials, altermagnets (AM), has emerged, challenging this traditional understanding of magnetic crystals, and introducing a platform for novel technology (*2*).

Unlike AFMs, with sublattices connected by translation or inversion, the sublattices in AMs can be connected by rotation or by transformations involving both a rotation and a translation. Interest in altermagnets stems from their ability to exhibit properties characteristic of FMs despite their net zero magnetic moment, a combination previously believed to be incompatible. As a result, AMs have an alternating, anisotropic spin splitting, enabling the generation of spin polarized currents as in FMs, but without the need for a finite net magnetic moment. The AHE and various magnetooptical effects are also enabled by the same magneto-crystalline symmetry that allows the altermagnetic band splitting (*3–5*).

Those unique properties of AMs originate from the local crystal environment of non-magnetic atoms within the unit cell (*6*). Altermagnetism is predicted in a broad range of materials. RuO$_2$, CrSb, and MnTe emerged as key platforms where it is sought and studied (*4*, *7–17*). The altermagnetic spin splitting hosted by those materials was confirmed by photoemission measurements (*4*, *7–11*, *15*). In addition to this splitting, the AHE was experimental observed in AMs that possess specific orientations of the Néel vector, $\vec{L}=\vec{m}_1-\vec{m}_2$ ($\vec{m}_{1,2}$ are the individual sublattice magnetization vectors) (*6*, *13*). This AHE is an important marker of the magneto-crystalline symmetry of AMs and can be harnessed for the read-out of their magnetic state in non-volatile memory devices.

The development of materials with a strong and tunable AHE has become a significant focus of research in spintronics (*13*, *18–22*). Understanding and controlling ferromagnetic materials has historically facilitated advancements in magnetic memory and spintronics technology (*23*, *24*). While their magnetization is easy to sense and control, FM components are sensitive to stray external fields, which threatens device stability and limits scalability. The development of antiferromagnetic spintronics was motivated by the expectation that they would be robust against stray fields, have improved stability at nanoscales, and allow for faster switching (*25*). Developing AFM spintronics, is however, difficult because of challenges in electrically detecting and controlling the order parameter (*26*). AMs that potentially host a strong and tunable AHE can offer a crucial solution to these challenges enabling spintronic applications based on AFM spintronics. This potential motivates a deeper fundamental understanding of their electrical response (*27*).

α-MnTe is an altermagnetic semiconductor and is hence ideal to investigate and tune the electrical signatures of altermagnetism. It crystallizes in the NiAs structure shown in Fig. 1(A) with Mn and Te atoms occupying different layers along the [0001] direction. The three magnetic easy axes are oriented along the $\langle 1\bar{1}00 \rangle$ directions, and the crystal structure allows a weak canting magnetization $\vec{M} = M_z\hat{z}$ oriented along [0001] (*28*, *29*). Despite this nearly vanishing magnetization, MnTe exhibits a strong spontaneous anomalous Hall effect (AHE) allowed by symmetry when the Néel vector is aligned with one of the easy axes reported in (*28*, *30*, *31*). The observation of a spontaneous anomalous Hall resistivity reaching 0.1 µΩ.m to 1 µΩ.m is reported in both films and single crystals of MnTe (*5*, *13*, *29*, *32*). The interpretation of the AHE of MnTe on InP(111) is also related to specific interfacial defects or stoichiometries in some reports (*32*, *33*). Differences in the qualitative and quantitative behavior of the AHE in α-MnTe across the literature motivate experimental measurements studying its dependence on intrinsic material properties as well as on extrinsic tuning knobs.

In this work, we show that the AHE of α-MnTe thin films grown on GaAs(111) is unexpectedly "written and tuned on-demand" by cooling down the material under an in-plane magnetic field. The magnitude and the direction of the cooling field have a dramatic impact on the strength and qualitative shape of the AHE. It can effectively manipulate its size and coercivity. This tunable character is unexpected since it is not consistent with the picture of the symmetry allowed anomalous Hall effect in this material, even after we account for magnetic domain population. We find this behavior to be unique to α-MnTe grown on GaAs and by extension other III-V materials, as it is absent in α-MnTe grown on SrF$_2$. There are subtle structural and morphological differences between α-MnTe grown on GaAs and α-MnTe grown on SrF$_2$ that correlate with this behavior. Notably, α-MnTe films grown on GaAs are thermally strained and possess a reduced c/a ratio compared to α-MnTe on SrF$_2$ and bulk which may significantly change the electronic structure of the material. Our results demonstrate that the AHE of strained MnTe grown on III-V semiconductors may be unique and fundamentally different from pristine α-MnTe. Its tunable character can lay the foundation for designing spintronic devices using altermagnets.

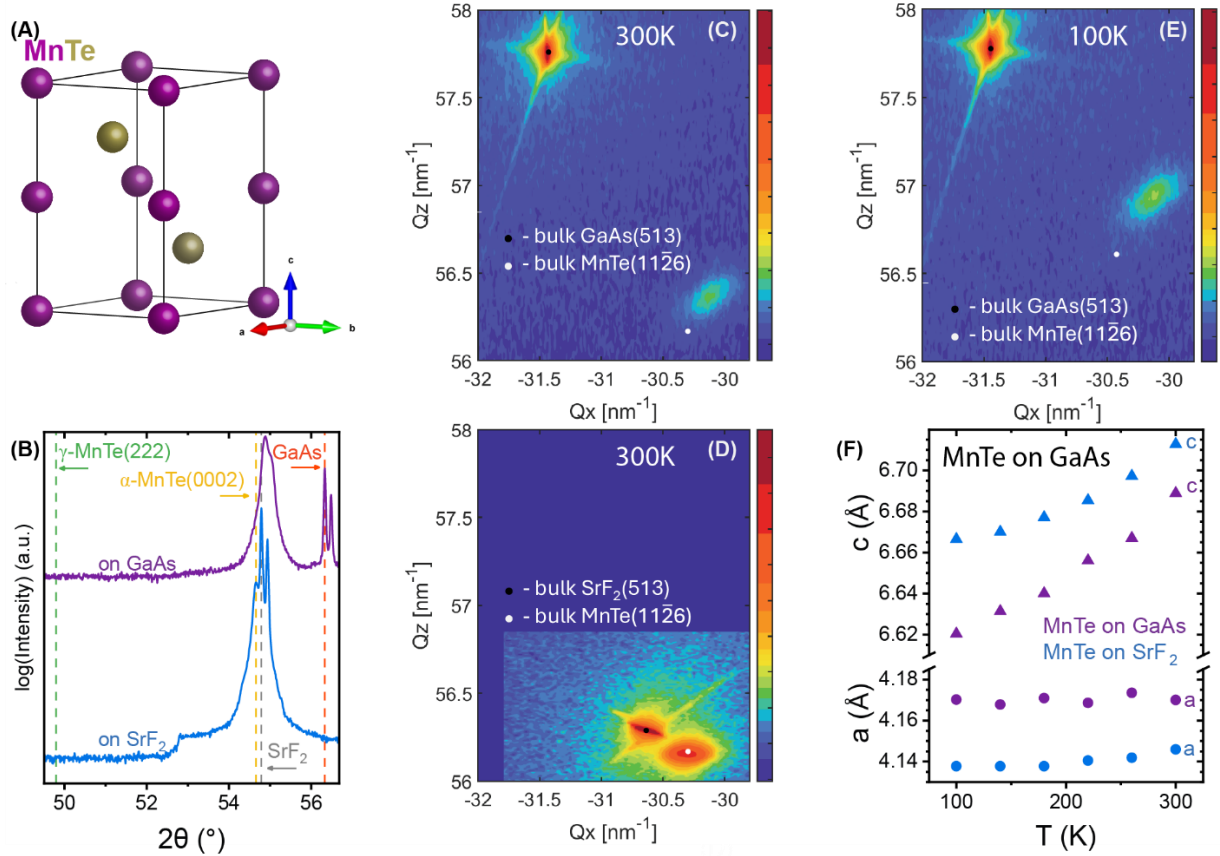

**FIG 1. Characterization of the crystal structure of MnTe grown on GaAs(111) and SrF$_2$(111) substrates.**
(A) Crystal structure of MnTe. (B) Specular X-ray diffraction patterns taken near the (222) substrate peak. (C) Reciprocal space map of α-MnTe grown on GaAs(111) taken at 300K. (D) Reciprocal map of α-MnTe grown on SrF$_2$ (111) at 300K. (E) Reciprocal space of α-MnTe on GaAs(111) taken at 100K. (F) Temperature dependence of the c and a lattice constants of α-MnTe grown on GaAs(111) and SrF$_2$.

**Structural characterization**

Hexagonal MnTe thin films are grown on GaAs(111) and SrF$_2$(111) following a procedure described in the methods section. X-ray diffraction measurements on films grown on both substrates reveal the nucleation of a pure hexagonal phase α-MnTe. Fig. 1(B) shows a specular X-ray diffraction pattern taken near the (222) peak of each respective substrate. It highlights the observation of the (0004) peak of α-MnTe and the absence of any peaks associated with zincblende γ-MnTe phase. It is however evidenced from Fig. 1(B) that the location of (0004) peak of α-MnTe is substrate dependent. α-MnTe grown on GaAs is under strain.

The large mismatch between the thermal expansion coefficients of MnTe and GaAs lead to a significant thermal strain as revealed by the reciprocal space maps shown in Fig 1(C-E). As the GaAs substrate is cooled down from the growth temperature to room temperature, it contracts slower than the MnTe layer, producing tensile stress on the MnTe layer. Fig. 1(C) shows a reciprocal space map (RSM) taken near the (513) peak of the GaAs substrate. The $(11\bar{2}6)$ Bragg peak of α-MnTe is located in the bottom right corner. Using knowledge of the bulk lattice parameters of α-MnTe, we specify the expected scattering vector locations ($Q_z$ and $Q_x$) of the $(11\bar{2}6)$ peak (white dot in Fig. 1C). From that comparison, we conclude that the film grown on GaAs(111) is under in-plane tensile strain despite the lattice spacing in α-MnTe being 3.6% larger than GaAs. Hence, this strain cannot be due to the lattice mismatch and must be a result of the mismatch between the thermal expansion coefficients of GaAs ($5.9 \times 10^{-6}\ K^{-1}$) and that of MnTe ($16.2 \times 10^{-6}\ K^{-1}$) near room temperature (*34, 35*).

We confirm this by carrying out temperature dependent X-ray diffraction measurements at low temperatures. The low temperature RSM is shown in Fig. 1(E). RSMs measured about the GaAs(513) peak taken between 100K and 300K are shown in Fig S1. They yield the "c" and "a" lattice parameters plotted versus temperature in Fig 1(F). The thermal expansion-induced strain observed in the film grown on GaAs(111) as a function of temperature and its changing c/a ratio are consistent with previous observations in MnTe films grown on InP(111) (*28*).

RSM measurements are also carried out on an α-MnTe film grown on lattice-matched SrF$_2$ (111) at room temperature. SrF$_2$ has a higher thermal expansion coefficient ($20 \times 10^{-6}\ K^{-1}$) than GaAs (*36*). The measurement yields an unstrained α-MnTe layer with lattice parameters consistent with bulk (see Fig. 1(D) and table 1). Apart from altering the in plane constant a and the out-of-plane lattice constant c, the thermal strain caused by the GaAs substrate changes the ratio c/a compared to bulk. A comparison is shown in Table 1. For both MnTe grown on GaAs(111) and on SrF$_2$ (111) $\phi - scan$ reveal an identical epitaxial arrangement $[11\bar{2}0]_{MnTe} \parallel [1\bar{1}0]_{substrate}$ (Fig. S2 and S3).

| Substrate | Thickness | a (Å) | c (Å) | c/a | Ref. |
|---|---|---|---|---|---|
| GaAs | 100 nm | 4.1783 | 6.6901 | 1.601 | This work (300K) |
| GaAs | 100 nm | 4.1702 | 6.6204 | 1.588 | This work (100K) |
| InP | 50 nm | 4.174 | 6.682 | 1.600 | Ref. (*28*) (300K) |
| SrF$_2$ | 145 nm | 4.1459 | 6.713 | 1.619 | This work (300K) |
| SrF$_2$ | 145 nm | 4.1377 | 6.666 | 1.610 | This work (300K) |
| SrF$_2$ | 2000 nm | 4.1485 | 6.718 | 1.619 | Ref. (*28*) (>300K) |
| None | Bulk | 4.1479 | 6.7118 | 1.618 | Ref. (*34*) (300K) |

Table 1. Structural parameters of α-MnTe grown on various substrates compared to bulk. Complete temperature dependent X-ray diffraction data is shown in the supplement.

**Tuning the AHE of α-MnTe through field cooling.**

To manipulate and measure the AHE of α-MnTe we employ a field cooling scheme illustrated in the insets of Fig. 2. We initially heat the sample to 340K, above its Néel temperature T$_N$. We then cool it down to the base temperature of cryostat (2K) under an applied magnetic field **B**$_{FC}$ (FC stands for field-cooled). **B**$_{FC}$ is always parallel to the basal plane of the crystal but its orientation in the plane is varied between different field cooling schemes. After reaching the base temperature, the magnetic field is removed and the sample is rotated to measure the AHE under an applied magnetic field **B**$_{Ex}$ ∥ [0001]. The sequence is repeated for different strengths of **B**$_{FC}$, and for orientations either along ($B_{FC} \parallel [1\bar{1}00]_{MnTe}$) or perpendicular to the magnetic easy axes ($B_{FC} \parallel [11\bar{2}0]_{MnTe}$). We measure the transverse (Hall) and longitudinal resistivity ($\rho_{xy}$ and $\rho_{xx}$) at each temperature by sweeping a magnetic field **B**$_{ex}$ applied perpendicular to the basal plane. A sizeable positive slope indicates that the ordinary Hall effect dominates the transverse resistivity and allows us to extract the carrier density (see Fig. S4, S5 and Table S1). We remove this contribution by subtracting a linear slope fit at high fields and we plot the anomalous part of $\rho_{xy}$. Unexpectedly, this field cooling scheme allows us to manipulate the AHE of α-MnTe grown on GaAs(111).

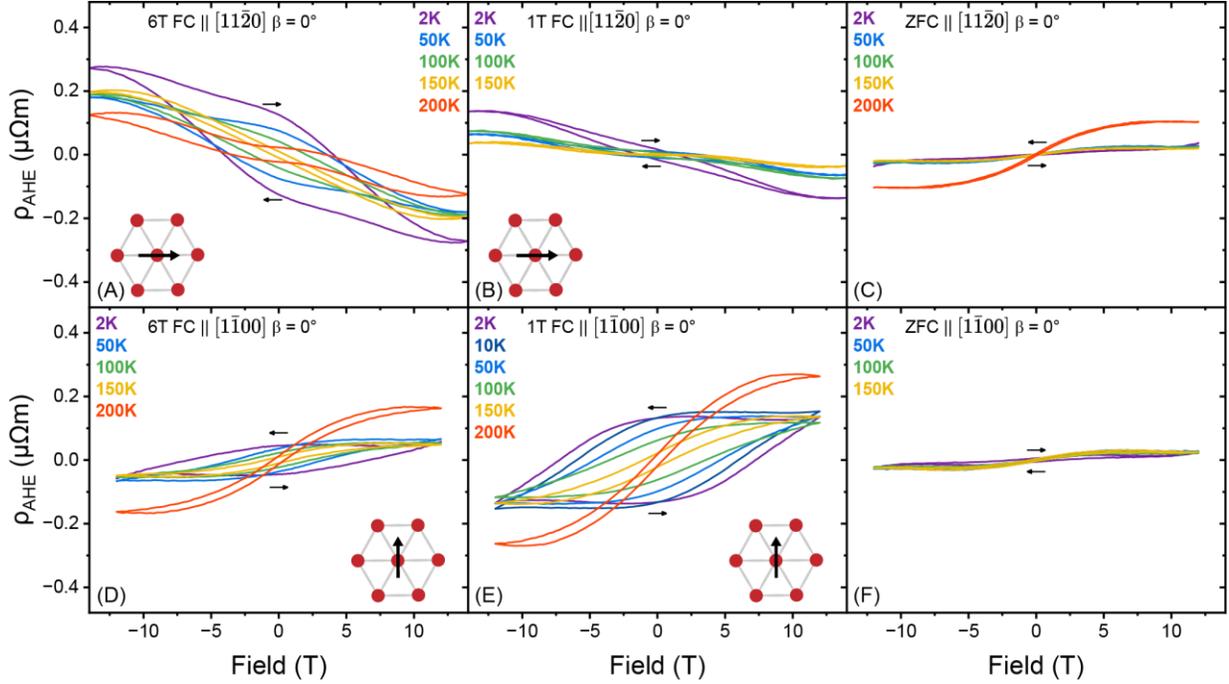

**FIG 2. Tuning the AHE of α-MnTe on GaAs with field cooling (FC).** Anomalous Hall resistivity $\rho_{AHE}$ for a 300μm wide Hall Bar of 100nm thick α-MnTe on GaAs(111) after removing the ordinary Hall contribution. $\rho_{AHE}$ plotted at different temperatures for (A) $B_{FC}$ = 6T || $[11\bar{2}0]$, (B) $B_{FC}$ = 1T || $[11\bar{2}0]$, (C) zero field cooling (ZFC), (D) $B_{FC}$ = 6T ||$[1\bar{1}00]$, (E) $B_{FC}$ = 1T ||$[1\bar{1}00]$, and (F) ZFC. **I** || $[1\bar{1}00]$ and $B_{EX}$ || $[0001]$ for all measurements. Arrows indicate the hysteresis direction in all frames. Inset represents the direction of $B_{FC}$ within the c-plane. β is the angle between **B** and the basal plane, where β=90° when **B** is along the c-axis.

Figure 2(A,B) shows the AHE of a 100nm α-MnTe film grown on GaAs(111) after the sample is cooled-down under $B_{FC}$ || $[11\bar{2}0]_{MnTe}$ equal to 6T and 1T respectively. We observe a spontaneous AHE for both values of $B_{FC}$ up to T=150K. For $B_{FC}$ = 6T, the magnitude of the spontaneous AHE as well as its coercive field are significantly enhanced. The AHE also exhibits a minor loop behavior and qualitatively suggests the presence of two AHE components. A hard component can be seen only below 100K and yields a coercive field as high as 4T at 2K. A softer component can be seen up to 150K and saturates to a higher value at 14T. The hard component is absent for $B_{FC}$ = 1T, but the spontaneous AHE is still visible. The minor loop behavior and the large coercive field are qualitatively similar to the AHE observed in Ref. (*33*). In Fig. 2(C), we plot the anomalous Hall resistivity after cooling the sample from $T_N$ in the absence of a magnetic field. The spontaneous Hall signal vanishes for zero field cooling (ZFC) confirming that field cooling activates the AHE on-demand.

In Fig. 2(D,E), we repeat the same field cooling process with $B_{FC}$ || $[1\bar{1}00]_{MnTe}$. We plot the anomalous Hall resistivity obtained after the sample is cooled-down under $B_{FC}$ equal to 6T and 1T, respectively. Again, we observe a spontaneous AHE up to T=150K. The anomalous Hall resistivity in Fig. 2(D,E) is positive and exhibits a minor loop behavior for both $B_{FC}$=6T and 1T. Unlike in Fig. 2(A,B), where a larger $B_{FC}$ enhances the AHE, Fig. 2(D,E) shows that field cooling along the $[1\bar{1}00]$ easy axis yields a stronger signal for $B_{FC}$=1T. In Fig. 2(F), we plot the results of the anomalous Hall resistivity for zero field cooling for this second sample geometry. We do not observe any detectable spontaneous AHE after ZFC, similar to the results of Fig. 2(C) confirming our previous observation. The combined results shown in Fig. 2(A-F) demonstrate that field

cooling is essential to controlling the AHE, and its success largely depends on the direction and magnitude of $B_{FC}$

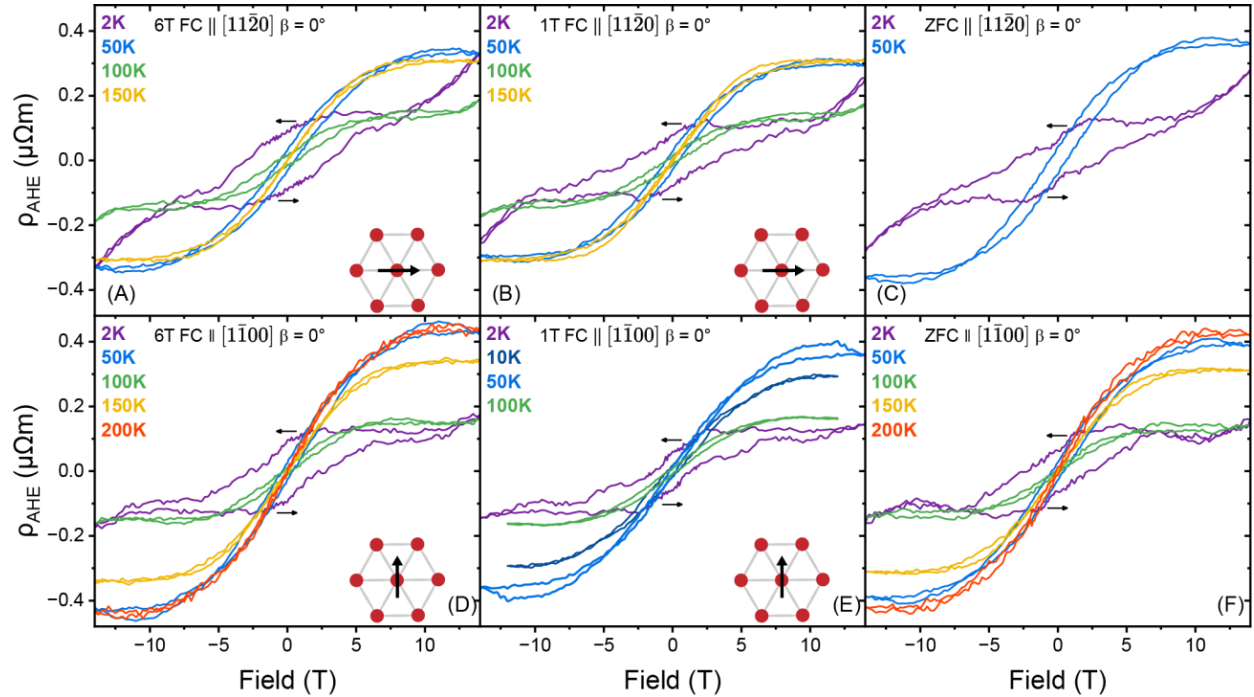

**FIG 3. AHE of α-MnTe on SrF$_2$.** Anomalous Hall resistivity of a rectangular piece cleaved from a 145nm thick α-MnTe film grown on SrF$_2$ meaured in a five-wire geometry. $\rho_{AHE}$ plotted at different temperatures for (A) $B_{FC}$ = 6T || $[11\bar{2}0]$, (B) $B_{FC}$ = 1T || $[11\bar{2}0]$ (C) zero field cooling (ZFC), (D) $B_{FC}$ = 6T || $[1\bar{1}00]$ (E) $B_{FC}$ = 1T || $[1\bar{1}00]$ and (F) ZFC. **I** || $[1\bar{1}00]$ and $B_{EX}$ || $[0001]$ for all measurements. Arrows indicate the hysteresis direction in all frames. Inset represents the direction of $B_{FC}$ within the c-plane.

The AHE and its tuning by field cooling shown in Fig. 2 are unique to the samples grown on GaAs(111) and are possibly features of thermally strained α-MnTe grown on III-V semiconductors (GaAs and InP). In Fig. 3, we repeat the same field cooling protocol for α-MnTe grown on SrF$_2$(111). In Fig. 3(A,B), we plot the anomalous Hall resistivity obtained after the sample is cooled down under $B_{FC}$ || $[11\bar{2}0]_{MnTe}$ equal to 6T and 1T, respectively. A prominent hysteresis loop is observed at 2K for both $B_{FC}$=1T and $B_{FC}$=6T. It decreases quickly with increasing temperature, completely vanishes by 150K. An identical behavior is observed when α-MnTe grown on SrF$_2$ is zero-field cooled (Fig. 3(C,F)) and when it is field cooled under $B_{FC}$ || $[1\bar{1}00]_{MnTe}$ (Fig. 3(D,E)).

The results presented in Fig. 3(A-F) for α-MnTe on SrF$_2$(111) show a different picture than those plotted in Fig. 3(A-F) for MnTe on GaAs(111). While a spontaneous AHE is visible for MnTe grown on SrF$_2$, it is insensitive to the sample cooling history. We discuss possible origins of the AHE in MnTe and its tuning in light of those findings.

**Origin of the AHE.**

Note here the AHE in collinear ferromagnets is proportional to the magnetization, and the hysteresis loops measured through $\rho_{AHE}(H)$ and $M(H)$ are qualitatively identical. Our results clearly show that this traditional picture is insufficient to explain the observed AHE, in both MnTe grown on GaAs and on SrF$_2$. In MnTe on GaAs, we do measure a weak remanent magnetization at low temperature much smaller than $10^{-2}$ μ$_B$/Mn (Fig. S6). It is largely independent of field cooling from 340K. Its hysteresis behavior is significantly softer than the hysteresis of the AHE. Despite this small magnetization, the spontaneous Hall resistivity reaching 0.1μΩ.m at 2K after field cooling, comparable to previous reports of the AHE in α-MnTe thin films (*13*). The magnetization observed for MnTe on SrF$_2$ is the expected response for an antiferromagnet, with a spin flop transition occurring between 0.5-1T ((Fig. S6)) (*5*, *37*). The spontaneous Hall resistivity at 2K also reaches 0.09μΩ.m. In both cases, the lack of proportionality between the AHE and the magnetization indicates its deviation from the conventional origin of the AHE, and further reveals the symmetry characteristic of MnTe films enabling its altermagnetic character.

According to previous work, the leading order spontaneous Hall conductivity allowed by symmetry in α-MnTe is cubic with the Néel vector (*38*):

$$\sigma_{xy} = a_3 L_y \left(3 L_x^2 - L_y^2\right) \quad (1)$$

Here $L_{x,y}$ are the in-plane (x,y) coordinates of the Néel vector $L$, where the y-direction is taken to be along a magnetic easy axis (see Fig. 4).

When multiple magnetic domains are present in real materials, the net Hall conductivity over all domains determines the spontaneous AHE. Previous investigations of MnTe thin films have shown via neutron diffraction and anisotropic magnetoresistance (AMR) that it has the easy-plane magnetic anisotropy shown in Fig. 4(A) with six equivalent easy axis orientations. Those investigations have also demonstrated how field cooling (FC) tunes domain population (*28*, *37*). There are possible equivalent easy axes of MnTe (blue, pink and green arrows), related to each other by a C$_3$ rotation, along with their time reversed partners (red, yellow, and purple). In its pristine state, a large slab of MnTe should host a random distribution of these six domains. Field cooling in-plane, can favor the nucleation of one or more domains among the three C$_3$-symmetric domains. The applications of an out-of-plane magnetic field can break the symmetry between the time-reversed domains.

Our field cooling scheme can thus select between three situations that favor the nucleation of one, two or three majority domains depending on direction and strength of B$_{FC}$ (*28*, *37*). First, by cooling under $B_{FC} \parallel [11\bar{2}0]$, we favor the nucleation of two magnetic domains oriented perpendicular to this direction (red and blue). Then once we apply $B_{EX} = 14T \parallel [0001]$, we favor the formation of a single majority domain oriented along one of those two equivalent directions. This is the first situation shown in Fig. 4(B). Second, by cooling down at $B_{FC} \parallel [\bar{1}100]$, we favor the nucleation of domains that have a Néel vector as close as possible to the direction perpendicular to the chosen $[\bar{1}100]$ (Fig. 4(C)). Then by applying $B_{EX} \parallel [0001]$, favor two out of those four domains. This is the second possible situation shown in Fig. 5(C). Lastly, when $B_{FC} = 0$ all six possible domains are equally present, and the application of $B_{EX} \parallel [0001]$ prior to the measurement favors three possible domains (Fig. 4(D)).

Let us now examine the consequences of this domain population change on Eq. (1). It is easy to see that the Hall conductivity is finite and equal for all three situations illustrated in Fig. (4), as long as the out-of-plane field identically breaks the symmetry between the time-reversed domains. The measurements of the AHE of α-MnTe on SrF$_2$ agree with this picture and is likely an intrinsic AHE. The tunable AHE of α-MnTe on GaAs violates Eq. (1) and must originate from other contributions.

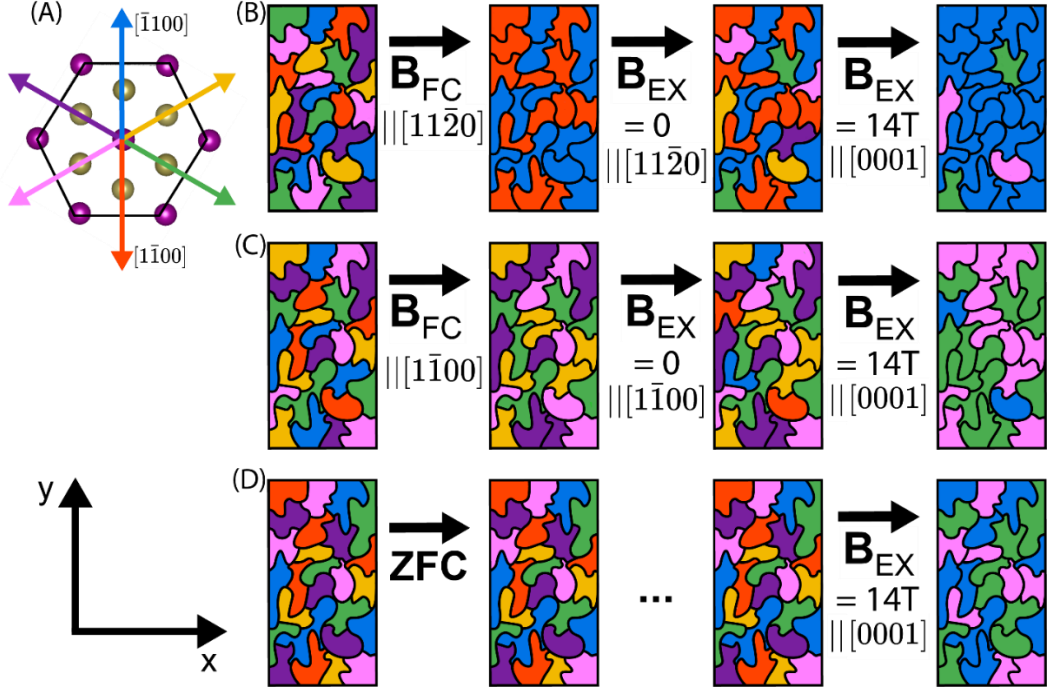

**FIG 4. (A)** Basal plane of MnTe showing six possible orientations of the Néel vector. Domain population after cooling MnTe following three different schemes: **(B)** $B_{FC} \parallel [11\bar{2}0]$ and applying $B_{EX} \parallel [0001]$, **(C)** $B_{FC} \parallel [1\bar{1}00]$ and applying $B_{EX} \parallel [0001]$ and **(D)** after zero field cooling and applying $B_{EX} \parallel [0001]$.

**Possible origin of the tuning.**

We propose two potential explanations for the unexpected tuning of the AHE observed in MnTe on GaAs. It is first possible that a coexistence of antiferromagnetism with other magnetic textures causes this hard and tunable AHE. Previous work has shown that large area films of MnTe grown on InP can host highly non-trivial two-dimensional magnetic textures that combine vortex-like regions with antiferromagnetism. while narrower micrometric channels of MnTe favor the nucleation of collinear antiferromagnetic domains (*39*) without vortices. On GaAs, MnTe grows as a uniform layer (Fig. S7). But on $SrF_2$, it forms a network of narrow channels with facets oriented along specific crystalline axes similar to the morphology of IV-VI materials grown by MBE (Fig. S8) (*40, 41*). It is thus possible that on $SrF_2$, MnTe predominantly hosts collinear antiferromagnetic domains favored by a channel-like morphology, which yields an AHE consistent with Eq. (1). Conversely, on GaAs, MnTe can host a complex planar magnetic texture, similar to the one revealed for large area samples in ref. (*39*) which would in turn yield a tunable AHE.

It is also possible that the tunable AHE is driven by a changing electronic or magnetic structure which can be caused by changing c/a ratio of α-MnTe on GaAs. It is in fact known that the c/a ratio alters the shape of the Fermi surface and shifts the valence band maximum to the Γ-point (*42*). A theoretical examination of the how Berry curvature depends of this changing band structure is needed. Note that the Berry curvature of MnTe exhibits a multipolar character in momentum space, as highlighted by (*5*) so it may be highly sensitive to the fine details of the band structure. Follow up studies of the Berry curvature of strained MnTe can shed light on the origin of the tunable AHE that we reveal. The c/a ratio can also alter the anisotropy energy (*43*). A weaker anisotropy can make domain manipulation more efficient favoring the observed tunable character of the AHE.

Lastly note that despite the larger carrier density seen in MnTe on GaAs, the saturation value of its AHE is weaker than what we find for MnTe on SrF$_2$. Under ZFC, the remanent AHE is also stronger in MnTe grown SrF$_2$. If the point defects (Mn$_{Te}$) which are responsible for the p-doping of MnTe were also responsible for the generation of an impurity related ferromagnetic AHE, one would expect this AHE to be larger when the impurity concentration is larger. Our data is not consistent with this picture.

In conclusion, our study provides critical insights into the highly sensitive nature of the AHE in α-MnTe, particularly in relation to substrate choice and cooldown history. We demonstrate that the AHE in MnTe on GaAs can be precisely tuned on-demand via field cooling, exhibiting behavior that deviates from predictions based on the symmetry-allowed Hall order parameter (Eq. (1)). A cooldown magnetic field as low as 0.03T is sufficient to achieve this tuning (Fig. S9), demonstrating its potential for applications. In contrast, the AHE observed in MnTe on SrF$_2$ aligns with the expected intrinsic AHE and is not tunable, underscoring the material's substrate-dependent electronic behavior. These findings highlight several key considerations for future research on this altermagnetic semiconductor. The electronic structure of MnTe is exceptionally sensitive to strain, and its magnetic domain dynamics are significantly influenced by both geometrical factors and cooldown history. The emerging interest in altermagnets, particularly in the altermagnetic semiconductor MnTe, is strongly justified by its demonstrated potential for tunability, as evidenced by our work. This tunable characteristic holds significant promise for the development of next-generation antiferromagnetic spintronic devices, offering tremendous benefits in terms of precision control and functionality.

**Methods**

**MBE growth procedure.** Semiconducting films of MnTe are grown on GaAs(111)B and SrF$_2$(111) substrates under ultra-high vacuum by molecular-beam epitaxy. First, the substrates are annealed at high temperature inside of the growth chamber to remove any surface oxide. For GaAs, an additional tellurium treatment at a high substrate temperature improves the surface quality further. The growth is monitored in-situ via Reflection High-Energy Electron Diffraction (RHEED). 100nm thick MnTe is deposited on GaAs at a growth temperature of 400°C at a rate of roughly 0.24 Å/s. On SrF$_2$, 145nm thick MnTe is deposited at a growth temperature of 340°C at a rate of approximately 0.4 Å/s. After growth, the sample is brought to room temperature over the span of 45 mins ± 10 mins in vacuum.

**Hall bar fabrication.** We pattern a 300µm wide Hall bar on a 100nm thick film of MnTe grown on GaAs using photolithography. The film is coated with MICROPOSIT® S1813® positive photoresist, and the pattern is carefully aligned with a chosen crystallographic direction. Directly after exposure, the sample is developed using AZ 917 MIF developer. Wet chemical etching with a H$_3$PO$_4$:H$_2$O$_2$:H$_2$O solution of volume ratio 1:3:6 is used to remove the MnTe film. After etching the sample is rinsed in DI water and blown dry with nitrogen. Remaining photoresist is removed using acetone. Finally, the sample is rinsed in ethanol, and blown dry with nitrogen.

**X-ray diffraction at room temperature.** Specular XRD was measured using a Malvern Panalytical Aeris X-ray diffractometer equipped with a Cu K-α source. Samples are measured in Bragg-Brentano geometry between 5-110° 2θ. The reciprocal space map of α-MnTe grown on SrF$_2$ was measured using a Bruker D8 Discover diffractometer equipped with a Cu K-α source.

**Temperature dependent X-ray diffraction.** Reciprocal space maps (RSMs) of symmetric (111/002 GaAs/MnTe) and asymmetric (513/116 GaAs/MnTe) diffraction spots were collected from 100 K to 300 K on the GaAs/MnTe sample with step sizes of 40 K. For these measurements, a Bruker D8 discover diffractometer was utilized, equipped with a rotating anode Cu Kα source, a Ge (004) monochromator, a liquid nitrogen-cooled Anton Parr DCS500 cold XRD stage, and a Dectris Eiger 2R 500 K area detector. At each temperature, alignment was conducted using a predicted d-spacing for the (111) GaAs peak with the assumption of an isotropic coefficient of thermal expansion for the cubic structure (*35*).

**Atomic force microscopy.** All atomic force microscopy scans were measured using the Jupiter XR AFM equipped with an Opus 160AC series cantilever operating in standard tapping mode.

**Magnetotransport measurements.** Magnetotransport measurements were carried out using a rotating probe inserted into a Quantum Design Dynacool 14T PPMS. For all measurements, the current density was $j \sim 3 \times 10^5$ A/m$^2$. Excitation current and voltage detection both utilized the internal components of the PPMS system. The system allows us to generate magnetic fields up to 14T in a temperature range up to 400K and down to 1.8K. After changing temperatures, we wait at least 20 minutes to ensure a stable measurement temperature. All samples were measured on warming, after cooling from 340K.

**Magnetometry measurements.** SQUID magnetometry measurements were carried out in a Quantum Design MPMS XL magnetometer allowing magnetic fields up to 7T and temperature down to 2K.

**Transmission electron microscopy.** High-resolution STEM images were acquired using a double-tilt holder and an aberration-corrected Spectra 30-300 transmission electron microscope (Thermo Fisher Scientific, USA), equipped with a field emission electron gun and operated at 300 kV. High-angle annular dark field (HAADF) were collected using the Panther detection system (Thermo Fisher Scientific, USA). The cross-sectional TEM samples were prepared using a Helios G4 UX dual-beam system (Thermo Fisher Scientific, USA) with the standard lift-out technique.

**Supplementary materials:**
**The pdf file includes**
Fig. S1-S9
Table S1.

**Acknowledgements** SB is partly supported by DOE-BES award DE-SC-0024291. JW, BAA and XL also acknowledge funding from NSF-DMR-1905277 and NSF-DMR-2313441. The work at NRL was supported by the Office of Naval Research (ONR) through the Naval Research Laboratory's Basic Research Program. Part of this research was conducted (A.V.I.) at the Center for Nanophase Materials Sciences (CNMS), which is a DOE Office of Science User Facility.

# Supplementary material for Unexpected Tuning of the Anomalous Hall Effect in Altermagnetic MnTe thin films

Sara Bey[1], Shelby S. Fields[2], Nicholas G. Combs[2], Bence G. Márkus[1,3], Dávid Beke[1,3], Jiashu Wang[1], Anton V. Ievlev[4], Maksym Zhukovskyi[5], Tatyana Orlova[5], László Forró [1,3], Steven P. Bennett[2], Xinyu Liu[1], Badih A. Assaf[1]

[1] Department of Physics and Astronomy, University of Notre Dame, Notre Dame IN, 46556, USA.
[2] Materials Science and Technology Division, U.S. Naval Research Laboratory, Washington DC, 20375, USA
[3] Stavropoulos Center for Quantum Matter, University of Notre Dame, Notre Dame IN, 46556, USA.
[4] Center for Nanophase Materials Science, Oak Ridge National Laboratory, Oak Ridge, TN, 37831, USA
[5] Notre Dame Integrated Imaging Facility, University of Notre Dame, Notre Dame, IN, 46556, USA

**Low temperature reciprocal space maps of α-MnTe on GaAs and SrF$_2$**

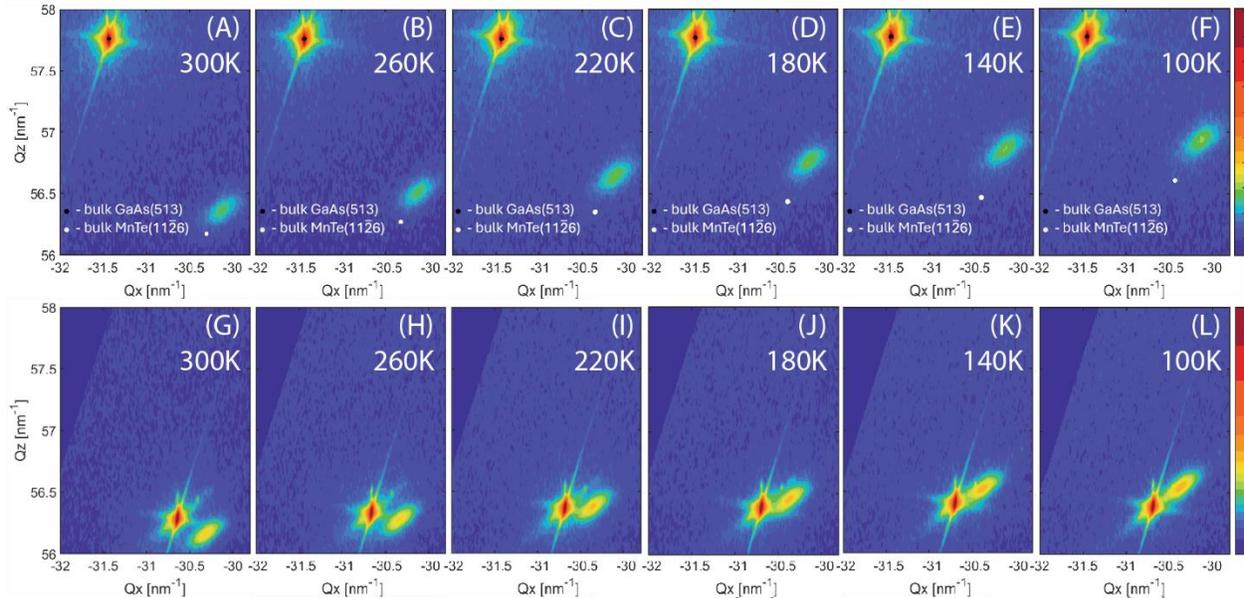

**Fig S1. RSM of the α-MnTe($11\bar{2}6$), GaAs(513), and SrF$_2$ Bragg peaks.** Measurements performed on warming at (A) 300K, (B) 260K, (C) 220K, (D) 180K, (E) 140K, and (F) 100K on 100nm thick α-MnTe grown on GaAs. Black and white dots represent the expected ($Q_x$,$Q_z$) position from knowledge of bulk lattice constants. Measurements on 145nm thick α-MnTe grown on SrF$_2$ performed on warming at (G) 300K, (H) 260K, (I) 220K, (J) 180K, (K) 140K, and (L) 100K. Substrate peak position shifted to known ($Q_x$,$Q_z$) position from knowledge of bulk lattice constants. (Ref. 14 and 35)

**Determination of the in-plane epitaxial relationship between the MnTe film and substrate**

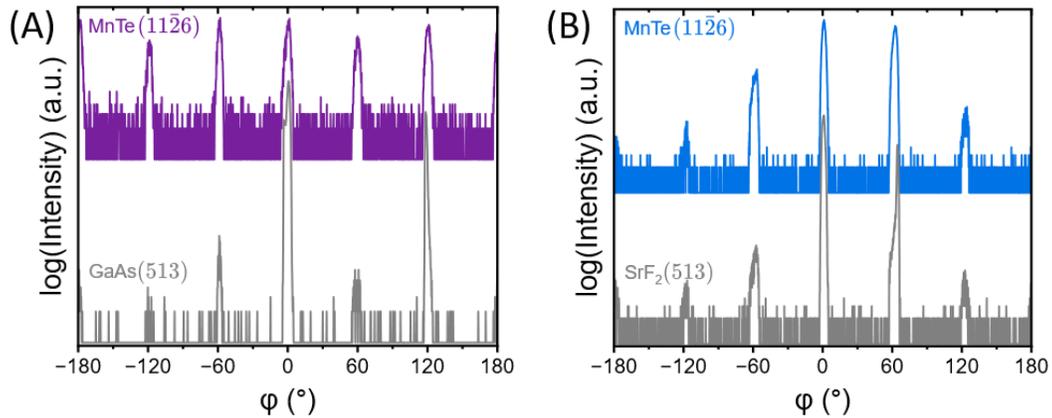

**Fig. S2. Phi-scan of asymmetric Bragg reflections for MnTe on GaAs and SrF$_2$** (A) 360° φ-scan of the asymmetric MnTe(11$\bar{2}$6) and GaAs(513) peaks. (B) 360° φ-scan of the asymmetric MnTe(11$\bar{2}$6) and SrF$_2$(513) peaks. Data has been shifted for visibility.

In addition to the reciprocal space maps reported in the main text, we measure 360° φ-scans for MnTe films on GaAs (A) and SrF$_2$ (B) (Fig. S1,S2). The periodicity of the asymmetric MnTe(11$\bar{2}$6) Bragg peaks reveals the $C_6$ symmetry in the basal plane of MnTe. We repeat the φ-scan, aligning to the asymmetric (513) plane of the substrate. Comparing between measurements aligned to the film and substrate indicates the high-quality epitaxial relationship between film and substrate. From this, we identify that the MnTe[11$\bar{2}$0] direction is parallel to Substrate[1$\bar{1}$0] direction. This relationship is further evident in the TEM images shown in Fig. S3.

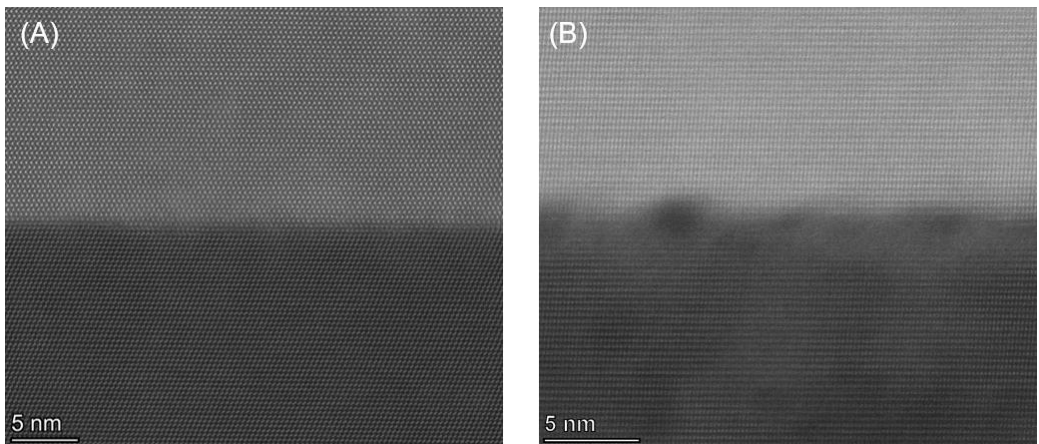

**Fig. S3. Cross-sectional scanning transmission electron microscope images.** Interface between the MnTe film and substrate for (A) MnTe on GaAs and (B) MnTe on SrF$_2$.

**Transverse resistivity showing combined ordinary Hall effect and AHE**

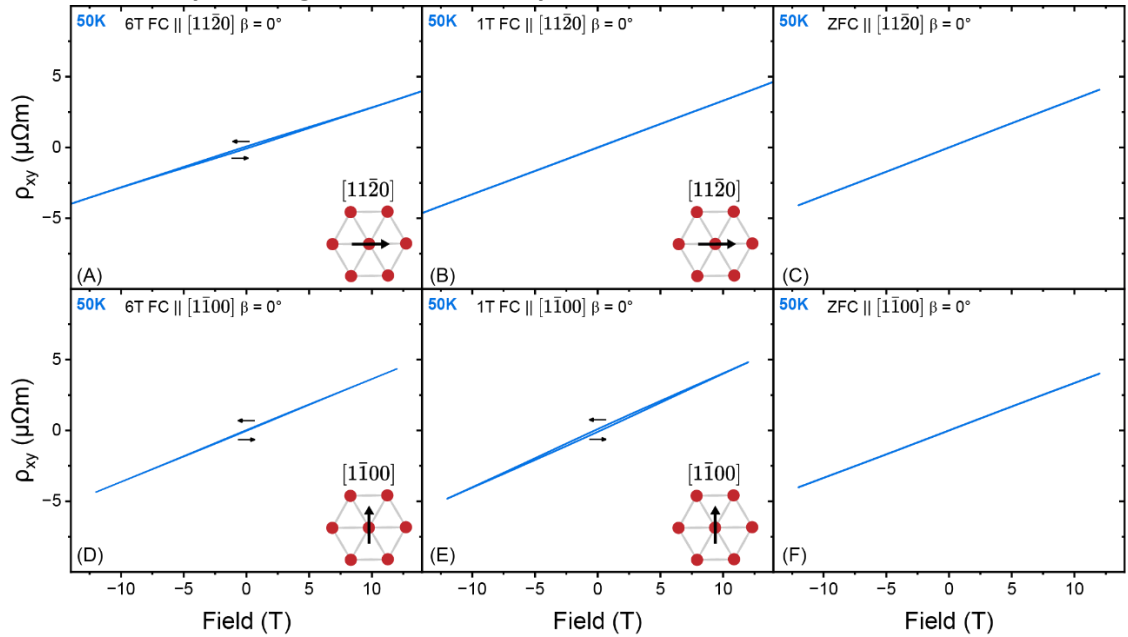

**Fig. S4 Hall resistivity for MnTe on GaAs(111) before subtracting the ordinary Hall effect.** Hall resistivity including both the ordinary and anomalous Hall contributions for a 300μm wide Hall Bar (HB) made on a 100nm thick α-MnTe on GaAs(111). **I** || $[1\bar{1}00]_{MnTe}$ and **B** || $[0001]_{MnTe}$ for all measurements. Arrows, inset, and β follow the same convention used in Fig. 3 and Fig. 4 of the main text. Total Hall resistivity at different temperatures for (A) $B_{FC}$ = 6T || $[11\bar{2}0]_{MnTe}$, (B) $B_{FC}$ = 1T || $[11\bar{2}0]_{MnTe}$, and (C) $B_{FC}$ = ZFC || $[11\bar{2}0]$. Total Hall resistivity at different temperatures for (D) $B_{FC}$ = 6T ||$[1\bar{1}00]_{MnTe}$, (E) $B_{FC}$ = 1T ||$[1\bar{1}00]_{MnTe}$, and (F) $B_{FC}$ = ZFC ||$[1\bar{1}00]_{MnTe}$.

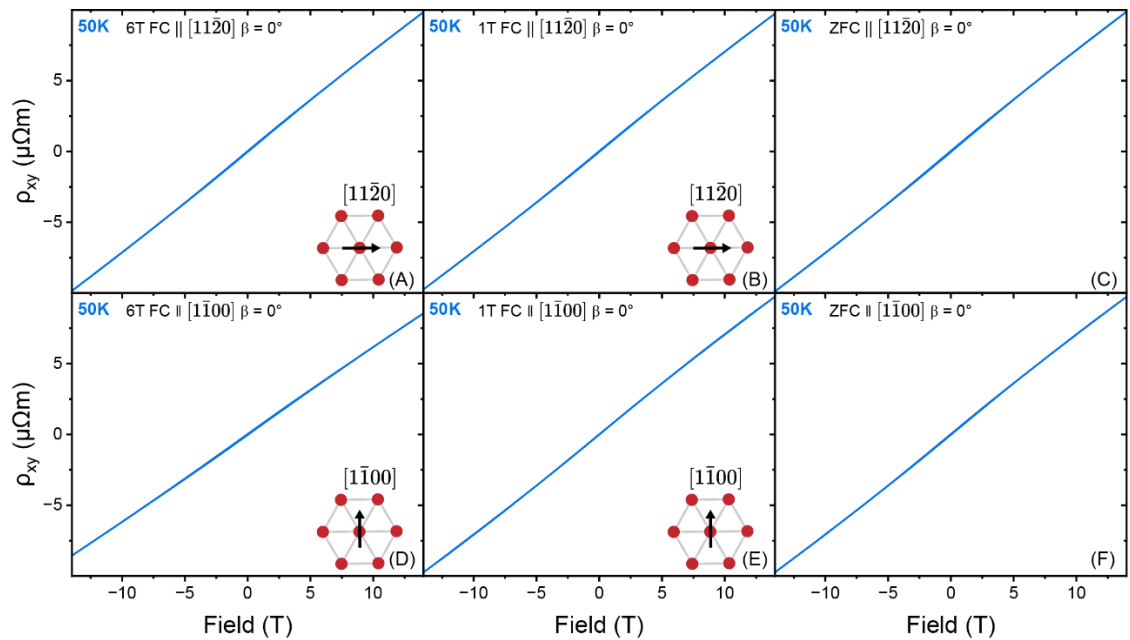

Fig. S5 **Hall resistivity for MnTe on SrF$_2$(111) before subtracting the ordinary Hall effect**. Hall resistivity including both the ordinary and anomalous Hall contributions for a rectangular piece cleaved from a 145nm thick α-MnTe film on grown on SrF$_2$. **I** || $[1\bar{1}00]_{MnTe}$ and **B** || $[0001]_{MnTe}$ for all measurements. Arrows, inset, and β follow the same

convention used in Fig. 3 and Fig. 4 of the main text. Total Hall resistivity at different temperatures for (A) $B_{FC}$ = 6T $||$ $[11\bar{2}0]_{MnTe}$, (B) $B_{FC}$ = 1T $||$ $[11\bar{2}0]$MnTe, and (C) $B_{FC}$ = ZFC $||$ $[11\bar{2}0]$. Total Hall resistivity at different temperatures for (D) $B_{FC}$ = 6T $||[1\bar{1}00]_{MnTe}$, (E) $B_{FC}$ = 1T $||[1\bar{1}00]_{MnTe}$, and (F) $B_{FC}$ = ZFC $||[1\bar{1}00]_{MnTe}$.

The total transverse (Hall) resistivity is shown for MnTe on GaAs in Fig. S4 and MnTe on SrF$_2$ in Fig. S5. These plots represent the signal before the ordinary contribution was subtracted, which corresponds to the anomalous Hall effect results presented in Fig. 2 and Fig. 3 of the main text, respectively. The raw, transverse resistivity is anti-symmetrized to eliminate any unintentional longitudinal resistivity signal due to slight misalignment of the Hall contacts. The anti-symmetrized transverse resistivity ρ$_{xy}$, in Fig. S4 and Fig. S5 is calculated as $\rho_{xy}(\downarrow) = \frac{\rho_{raw}(\downarrow)-\rho_{raw}(\uparrow)}{2}$ and $\rho_{xy}(\uparrow) = \frac{\rho_{raw}(\uparrow)-\rho_{raw}(\downarrow)}{2}$, where (↓) represents sweeping from +**B** to -**B** and (↑) represents sweeping from -**B** to +**B**. The ordinary Hall contribution is dominant in all cases, evidenced by the positive linear slope. Despite this, the hysteresis is visible before subtracting a slope at high field in both Fig. S4(A) and Fig. S4(E). These correspond to the strongest hysteresis observed in the anomalous Hall effect after tuning with the field cooling schemes reported in the main text.

The transverse resistivity for MnTe on SrF$_2$, shown in Fig. S5, shows the insensitivity of these samples to field cooling schemes. The signal originating from the ordinary Hall contribution is dominant for all field cooling strengths and directions in MnTe on SrF$_2$, indicated by the positive slope of the signal in Fig. S5(A-F). This is largely irrespective of the field cooling scheme, consistent with the results of the AHE presented in Fig. 3 of the main text.

**Carrier concentration, resistivity, and mobility calculated from magnetotransport at 50K**

| Sample | $B_{FC}$ | $|p|$ (cm$^{-3}$) | $\rho_{xx}$ (μΩm) | μ (cm$^2$V$^{-1}$s$^{-1}$) |
|---|---|---|---|---|
| GaAs HB | 6T $||$ $[11\bar{2}0]$ | 2.1 x 10$^{19}$ | 47.6 | 62 |
| GaAs HB | 1T $||$ $[11\bar{2}0]$ | 1.86 x 10$^{19}$ | 47.8 | 70 |
| GaAs HB | ZFC $||$ $[11\bar{2}0]$ | 1.85 x 10$^{19}$ | 47.7 | 71 |
| GaAs HB | 6T $||$ $[1\bar{1}00]$ | 1.75 x 10$^{19}$ | 46.6 | 77 |
| GaAs HB | 1T $||$ $[1\bar{1}00]$ | 1.6 x 10$^{19}$ | 46.6 | 84 |
| GaAs HB | ZFC $||$ $[1\bar{1}00]$ | 1.88 x 10$^{19}$ | 46.2 | 72 |
| GaAs R | 1T $||$ $[1\bar{1}00]$ | 3.9 x 10$^{19}$ | 43.3 | 37 |
| SrF$_2$ R | 6T $||$ $[11\bar{2}0]$ | 9.12 x 10$^{18}$ | 324 | 21 |
| SrF$_2$ R | 1T $||$ $[11\bar{2}0]$ | 9.19 x 10$^{18}$ | 321 | 21 |
| SrF$_2$ R | ZFC $||$ $[11\bar{2}0]$ | 9.12 x 10$^{18}$ | 324 | 21 |
| SrF$_2$ R | 6T $||$ $[1\bar{1}00]$ | 9.33 x 10$^{18}$ | 307 | 22 |
| SrF$_2$ R | 1T $||$ $[1\bar{1}00]$ | 9.25 x 10$^{18}$ | 307 | 22 |
| SrF$_2$ R | ZFC $||$ $[1\bar{1}00]$ | 9.29 x 10$^{18}$ | 308 | 22 |

**Table S1. Sample characteristics at 50K for MnTe on GaAs and SrF$_2$. HB: Hall bar. R: rectangle. FC: Field cooling, ZFC: zero field cooling.**

The transport parameters extracted from magnetotransport data at 50K are reported in table S1 for all field cooling protocols for MnTe on GaAs and SrF$_2$. The charge carrier density is calculated from the positive Hall slope using the simplified, single parabolic band picture, as $R_H = \frac{1}{e|p|}$, where R$_H$ is the Hall coefficient, e is the charge of the electron, and |p| is the carrier concentration. This provides an estimation of the concentration of holes in these films, since it neglects the complex valence band structure of MnTe (ref. 42). In light of this, we see that the calculated carrier concentration is a factor of two larger in MnTe on GaAs. The longitudinal resistivity of MnTe on SrF$_2$ is approximately six times larger than the resistivity of MnTe on GaAs.

**Negligible magnetization in MnTe on GaAs and SrF$_2$**

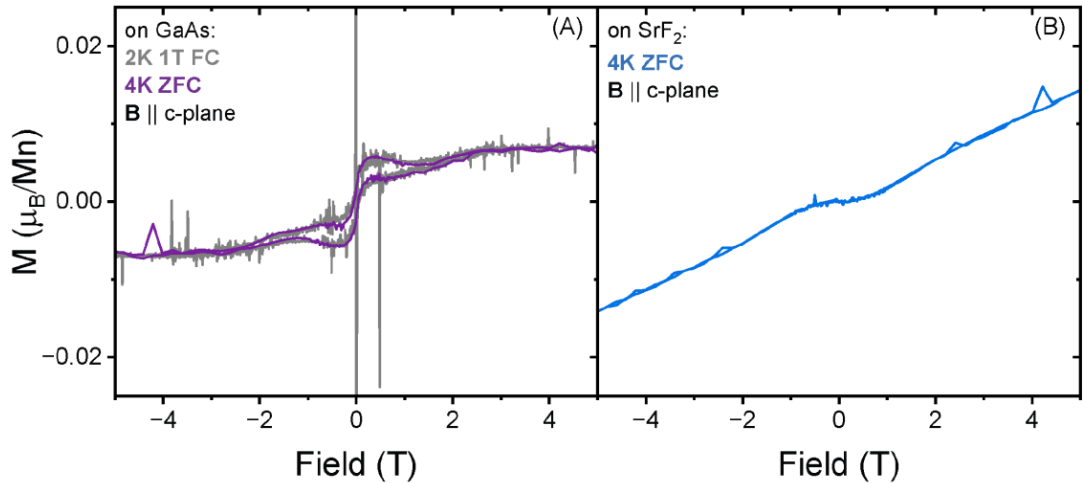

**Fig. S6. In-plane magnetization after subtracting diamagnetic background.** Magnetization at low temperature for 600nm MnTe on GaAs and 1080nm nm MnTe on GaAs. A slope is fit at high field, and subtracted to eliminate the diamagnetism from the substrate. (A) weak ferromagnetic magnetization response of MnTe on GaAs at low temperature for 1T FC from 340K and ZFC, and (B) antiferromagnetic magnetization response of MnTe on SrF$_2$ at 4K. In all measurements, the sample is positioned with the field applied in the basal plane.

In-plane SQUID magnetometry is measured with a Quantum Design MPMS system on a 600nm MnTe film grown on GaAs and on a 1080nm film grown on SrF$_2$. Due to the vanishingly small magnetization, the diamagnetic signal from the substrate dominates, and measurements of thinner films are challenging. A slope fit at high fields is subtracted to remove the diamagnetism of the substrate from the raw data and reveal the remaining signal of the film. In Fig. S6(A), the in-plane magnetization is shown for MnTe on GaAs after 1T FC and ZFC. The remaining signal reveals a remanent magnetization of < 0.01 $\mu_B$/Mn with a weakly ferromagnetic behavior. In addition, the weak signal for MnTe on GaAs is largely independent of field cooling from 340K. The magnetization observed for MnTe on SrF$_2$ (Fig. S6(b)) is the expected response for an antiferromagnet, with a spin flop transition occurring between 0.5-1T (ref. 37).

**Surface analysis via Atomic Force Microscopy (AFM)**

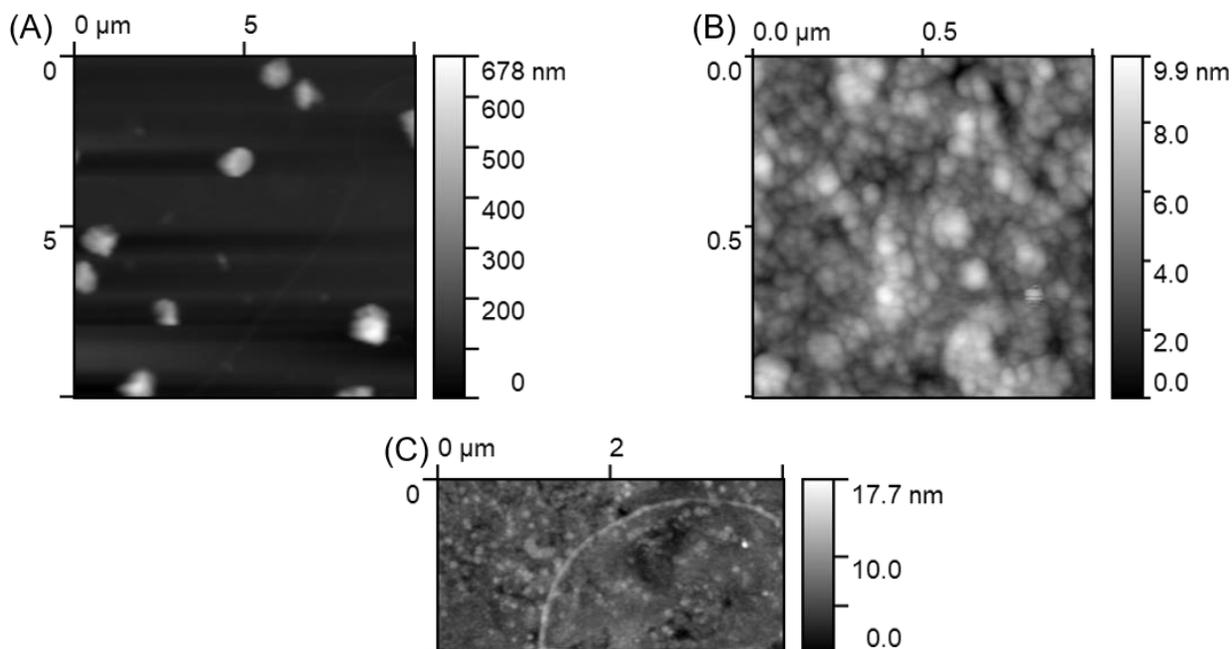

**Fig. S7. Morphology of α-MnTe on GaAs** (A) 10μm x 10μm AFM scan for 160nm MnTe on GaAs. Large features on surface have been identified as In and GaAs particles via SEM/EDS resulting from substate mounting and cleaving. The film is continuous. (B) 1μm x 1μm AFM scan away from contaminants reveals the an overall continous surface, with a root-mean-square (RMS) surface roughness of 1.44 nm, calculated for this 1μm² region. (C) 2μm x 4 μm AFM scan with a RMS surface roughness of 1.54nm.

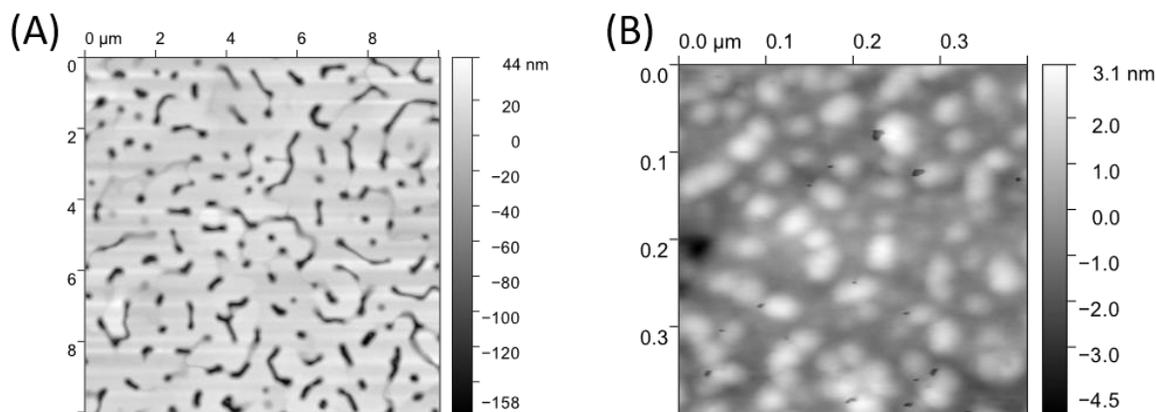

**Fig. S8. Morphology of α-MnTe on SrF₂** (A) 10μm x 10μm AFM scan for 145nm MnTe on SrF$_2$. Plateaus appear on the surface, with moderately deep seperations between them. The plateaus do not fully coalesce forming boundaries that oriented along hexagonal crystalline facets. (B) 0.4μm x 0.4μm AFM scan taken on a plateau region. The microscopic morphology appears smooth, with a RMS surface roughness of 1.0nm for this 1μm² region.

The MnTe film grown on GaAs(111) is continuous as can be seen in the Fig. S7(B,C). The topography of a 1μm² scan done on MnTe grown on GaAs(111) is shown in Fig S7(D). At this decreased scale, the contrast of the film surface is exposed, and we measure a RMS roughness, $S_q$=1.44 nm.

The morphology, shown in Fig. S8(A) for a 10 μm² region, presents features that are unique to MnTe on SrF$_2$(111), contrasting the results of Fig. S7 for MnTe on GaAs. This terrace morphology is similar to the surface of IV-VI materials grown on BaF$_2$ (ref. 40,41). The film is not fully coalesced, and clear divisions exist between terraces. The scan size is further decreased to 0.4 µm² in Fig. S8(B), to investigate the film roughness of an individual terrace, and the calculated RMS roughness is $S_q=1.0$ nm.

**Low magnetic field tuning of the AHE**

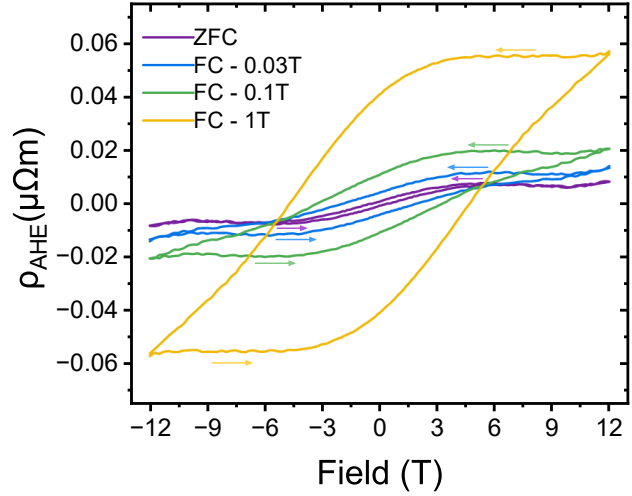

**FIG S9. Tuning the AHE of α-MnTe on GaAs with field cooling (FC).** Anomalous Hall resistivity $\rho_{AHE}$ for a rectangular piece of a 150nm thick α-MnTe on GaAs(111), after removing the ordinary Hall contribution. The Hall effect is measured with B applied along the c-axis. The cooldown field is applied in the c-plane along $[1\bar{1}00]$. The current is applied along $[1\bar{1}00]$. A cooldown field as low as 0.03T activates a spontaneous AHE.